\documentclass[conference]{IEEEtran}

\ifCLASSINFOpdf
\else
\fi

\usepackage{lineno}
\usepackage[
top    = 1.905cm,
bottom = 2.54cm,
left   = 2.00cm,
right  = 1.5748cm]{geometry}

\usepackage{setspace, amsmath, amssymb, url, lscape, subfigure, multirow, pslatex, listings, verbatim, alltt, amsfonts, wrapfig, boxedminipage, color, cite,float,algorithmic}
\usepackage[vlined,linesnumbered,ruled,boxed]{algorithm2e}
\usepackage{mathtools}

\usepackage{balance}

\usepackage[dvips]{graphicx}
\usepackage{epsfig}
\usepackage{amsmath}
\usepackage{xcolor}
\usepackage{mdframed}
\newcommand{\qed}{\nobreak \ifvmode \relax \else
      \ifdim\lastskip<1.5em \hskip-\lastskip
     \hskip1.5em plus0em minus0.5em \fi \nobreak
      \vrule height0.75em width0.5em depth0.25em\fi}

\newcommand{\eg}{{\it e.g., }}
\newcommand{\etal}{{\it et~al. }}
\newcommand{\ie}{{\it i.e., }}

\newcommand{\comments}[1]{}
\newcommand\hl{\bgroup\markoverwith
  {\textcolor{yellow}{\rule[-.5ex]{2pt}{2.5ex}}}\ULon}

\begin{document}

\title{Analyzing the Performance of Smart Industry 4.0 Applications on Cloud Computing Systems}

\author{
{\bfseries Razin Farhan Hussain$^1$, 
Alireza Pakravan$^2$, Mohsen Amini Salehi$^1$}\\
$^1$High Performance Cloud Computing (HPCC) Laboratory, University of Louisiana at Lafayette, LA, USA\\
$^2$California State University, San Marcos, CA, USA\\
{Email:$^1$\{razinfarhan.hussain1,amini\}@louisiana.edu, $^2$apakravan@csusm.edu}
}


\maketitle

\IEEEpeerreviewmaketitle
\begin{abstract}
Cloud-based Deep Neural Network (DNN) applications that make latency-sensitive inference are becoming an indispensable part of Industry 4.0. Due to the multi-tenancy and resource heterogeneity, both inherent to the cloud computing environments, the inference time of DNN-based applications are stochastic. Such stochasticity, if not captured, can potentially lead to low Quality of Service (QoS) or even a disaster in critical sectors, such as Oil and Gas industry. To make Industry 4.0 robust, solution architects and researchers need to understand the behavior of DNN-based applications and capture the stochasticity exists in their inference times. Accordingly, in this study, we provide a descriptive analysis of the inference time from two perspectives. First, we perform an application-centric analysis and statistically model the execution time of four categorically different DNN applications on both Amazon and Chameleon clouds. Second, we take a resource-centric approach and analyze a rate-based metric in form of Million Instruction Per Second (MIPS) for heterogeneous machines in the cloud. This non-parametric modeling, achieved via Jackknife and Bootstrap re-sampling methods, provides the confidence interval of MIPS for heterogeneous cloud machines. The findings of this research can be helpful for researchers and cloud solution architects to develop solutions that are robust against the stochastic nature of the inference time of DNN applications in the cloud and can offer a higher QoS to their users and avoid unintended outcomes.
\end{abstract}
\begin{IEEEkeywords}
Deep Neural Network Applications, Industry 4.0, Cloud Platform, Heterogeneous Machines, Inference Time
\end{IEEEkeywords}

\section{Introduction}\label{intro}
Software solutions operating based on machine learning and, particularly, Deep Neural Network (DNN) models are becoming fundamental pillars of Industry 4.0 revolution \cite{lu2019oil}. In the industrial automation process, numerous smart sensors frequently produce and fed data to the DNN-based applications that can make smart latency-sensitive decisions to improve energy efficiency, production, and safety measures. Building robust Industry 4.0 solutions entail having an accurate estimation of the inference (execution) time of DNN-based applications hosted on the cloud or edge computing systems. Lack of such assessments often leads to missing the applications' latency constraints and lowers their quality of service (QoS) \cite{li2017cost} or increase the incurred cost of cloud resources. In critical industrial sectors, such as oil and gas, the penalty of such inaccurate estimations can be disastrous and cause unintended consequences, such as an unsafe workplace, environmental footprints, energy wastage, and damaging devices \cite{hussain2019federated,dincer2015review}. Accordingly, our goal in this study is to measure and model the stochasticity that exists in the execution time of industrial DNN-based applications that are commonly deployed in the cloud.\looseness=-1 

Our motivation in this study is the critical industry of Oil and Gas (O\&G) that is aimed at becoming clean and ultimately unmanned, thereby safe, in Industry 4.0. O\&G is one of the main environmental pollutants and even minor improvements in this industry can have major impacts in the global scale. In this context, there are several time-sensitive operations (\eg fire detection \cite{dunnings2018experimentally} and toxic gas monitoring \cite{aliyu2016development}) that failing to timely process them can potentially lead to a disasters, scuh as oil spill, explosion, and even death. 
Understanding the uncertainties exist in execution time of different application types and properly modeling them is crucial in architecting software solutions that are robust against these uncertainties. Note that DNN-based applications encompass both the training and inference stages \cite{eshratifar2019jointdnn}. While the training stage is generally carried out offline, our focus in this study is on modeling the inference execution time that has to be accurately estimated for latency-sensitive and mission critical applications \cite{zhang2018efficient}. For instance, accurate estimation of inference time is instrumental in calculating the completion time of arriving tasks that can, in turn, help to make more precise resource allocation decisions \cite{han2019ondisc}. 

Public or private Cloud datacenters, such as Amazon \cite{bermudez2013exploring}, are widely used as the back-end platform to execute DNN-based industrial applications \cite{luckow2016deep}. The cloud providers often offer heterogeneous machine types, such as CPU-Optimized, Memory-Optimized, and GPU, that provide different execution time for various application types. For instance, a big data application type has its lowest execution time on the Memory-Optimized machine type whereas an image rendering application is best fitted to the GPU-based machine type. This form of heterogeneity is known as \textit{inconsistent heterogeneity} \cite{mokhtari2020autonomous,ipdps19}. For each machine type, cloud providers offer a \textit{consistent heterogeneity} in form of various virtual machine (VM) instance types with different number of allocated resources.
~For example, in Amazon cloud, for CPU-Optimized machine type, there is $c5d.18xlarge$ VM type with 36 number of cores that is faster than $c5.xlarge$ VM type with only 2 cores. As each application type can potentially have different inference time on distinct machine types, it is critical to consider the resource heterogeneity in estimating the execution time of different DNN-based applications.

For that purpose, in this study, we evaluate and analyze the execution time of DNN-based applications on heterogeneous cloud machine types. Our study encompasses both the application-centric perspective, by the way of modeling inference time, and the resource-centric perspective, by the way of measuring the Million Instruction Per Seconds (MIPS) metric. MIPS is considered as a rate-based metric that reflects the performance of cloud machine instance in terms of execution speed. As we consider latency-constrained applications, the underlying systems is considered as a dynamic (online) platform that processes each task upon arrival. 

Prior studies on evaluating and modeling DNN-based applications \cite{xia2019dnntune} mostly focus on the core DNN model and ignore the end-to-end latency of the application that includes at least two other factors: (a) the latency of non-DNN parts of the application (\eg those for pre- and post-processing); and (b) the latency imposed due to uncertainties inherent to the cloud platform. Nonetheless, for critical industrial applications, such as those in O\&G, a holistic analysis that considers the end-to-end latency of DNN-based applications is needed. The lack of such study hinders the path to develop a robust smart O\&G solutions \cite{hussain2018robust}. Accordingly, the main contributions of this work are as follows: 
\begin{itemize}
    \item Providing an application-centric analysis by developing a statistical model of the inference time of various DNN-based applications on heterogeneous cloud resources.
    \item Providing a resource-centric analysis of various DNN-based applications on heterogeneous cloud resources by developing a statistical model of MIPS, as a rate-based metric.
    \item Providing a publicly available\footnote{\label{note1}\url{https://github.com/hpcclab/Benchmarking-DNN-applications-industry4.0}} collection of pre-trained DNN-based industrial applications in addition to their training and testing datasets. Moreover, a trace of inference execution times of the considered applications on heterogeneous machines of two public cloud platforms (namely AWS and Chameleon Cloud \cite{chameleon}) is presented.
\end{itemize}

The rest of the paper is organized as follows: Section \ref{appType} discusses four different DNN applications and their underlying architectures. Section \ref{insType} states various cloud execution platforms with brief discussion. Section~\ref{expSetup} demonstrates the experimental setup to execute the DNN-based applications. The section~\ref{result} provides the application-centric analysis, whereas section~\ref{resCentric} presents the resource-centric analysis. Section~\ref{relatedwork} presents related works. Finally, section~\ref{concl} concludes the paper with discussion and future avenues for exploration.

\section{DNN-Based Applications in O\&G Industry 4.0}\label{appType}
Table \ref{table:apptype} summarizes different types of DNN-based applications used in the smart O\&G industry. The table shows the abbreviated name for each application, its DNN (network) model, type of its input data, the scope of deployment in O\&G Industry 4.0 \cite{nguyen2020systematic}, and the code base to build the model. All the applications, the input data, and analysis results are publicly available for reproducibility purposes in the Github repository mentioned earlier. In the rest of this section, we elaborate on the characteristics of each application type.

\begin{table}[h!]
\centering
\scalebox{0.62}
{\begin{tabular}{l||l|l|l|l}

\textbf{Application Type} & \textbf{DNN Model} & \textbf{Input Type} & \textbf{Scope} & \textbf{Code Base}\\ \hline \hline
\textit{Fire Detection (Fire)} & Customized Alexnet & Video Segment & \begin{tabular}[c]{@{}l@{}}Control \&\\ Monitoring\end{tabular} & \begin{tabular}[c]{@{}l@{}}Tensorflow \\ (tflearn)\end{tabular}\\ \hline
\textit{\begin{tabular}[c]{@{}l@{}}Human Activity\\ Recognition (HAR)\end{tabular}} & \begin{tabular}[c]{@{}l@{}}Customized Sequential \\ Neural Network\end{tabular} & Motion sensors & \begin{tabular}[c]{@{}l@{}}Workers \\ Safety\end{tabular} & keras\\ \hline
\textit{Oil Spill Detec. (Oil)} & FCN-8 & SAR Images & \begin{tabular}[c]{@{}l@{}}Disaster \\ Management\end{tabular} & keras\\ \hline
\textit{\begin{tabular}[c]{@{}l@{}}Acoustic Impedance \\ Estimation (AIE) \end{tabular}} & \begin{tabular}[c]{@{}l@{}}Temporal Convolutional\\ Network\end{tabular} & Seismic Data & \begin{tabular}[c]{@{}l@{}}Seismic\\ Exploration\end{tabular} & PyTorch\\ 
\end{tabular}}
\caption{\small{DNN-based applications used in O\&G Industry 4.0 along with their network model, input data type, usage scope, and code base.}}
\label{table:apptype}
\end{table}
\subsection{Fire Detection (abbreviated as Fire)}
Smart fire detection, a critical part of monitoring systems, aims at providing safety and robustness in Industry 4.0. We analyzed a
fire detection application developed by Dunnings and Breckon \cite{dunnings2018experimentally} using convolutional neural network (CNN). It automatically detects fire regions (pixels) in the frames of a surveilled video in a real-time manner. Among other implementations, we deploy the FireNet model that accurately identifies and locate fire in each frame of a given video segment. FireNet is a lightweight variation of AlexNet model \cite{alom2018history} with three convolutional layers of sizes 64, 128, and 256. In this model, each convolutional layer is augmented by a max-pooling layer and a local response normalization to achieve high frequency features with a large response from previous layer. 
To analyze the inference time of the fire detection system, we constructed a benchmarking dataset of 240 videos with different backgrounds. For fair and realistic analysis, the length of all videos is considered two seconds. 

\subsection{Human Activity Recognition  (abbreviated as HAR)}
Human Activity Recognition (HAR) systems are widely used in Industry 4.0 to ensure workers safety in hazardous zones. For this purpose, motion sensors, such as accelerometer and gyroscope, that are widely available on handheld PDA devices are utilized. The HAR system we use operates based on the sequential neural network model with four layers to identify the worker's activities (namely, walking, walking upstairs, walking downstairs, sitting). For analysis, we use a dataset of UCI machine learning repository, known as Human Activity Recognition Using Smartphones \cite{anguita2013public}. 
\begin{figure}[h!]
	\centering	
	\includegraphics[scale=0.45]{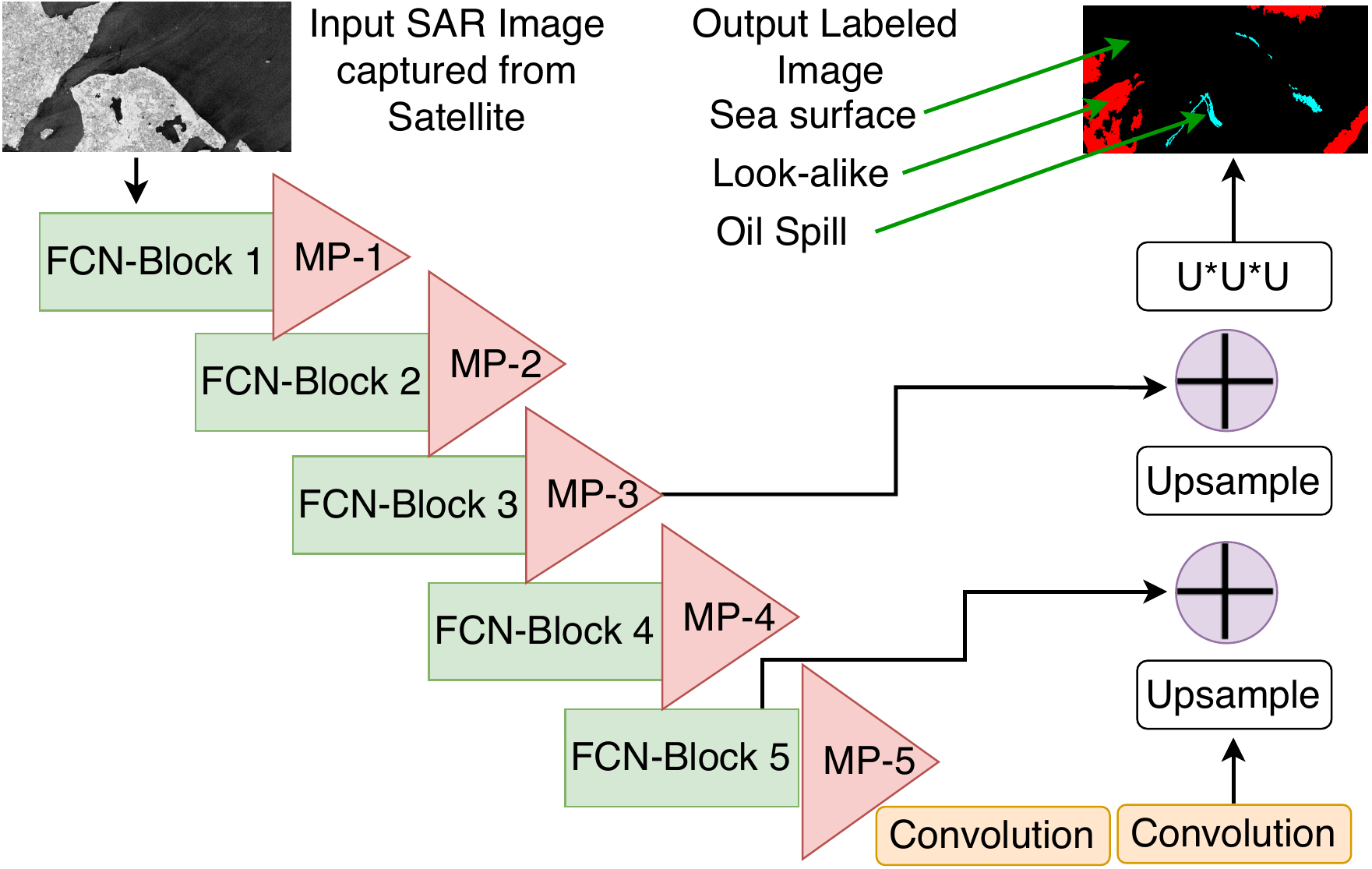}
	\caption{\small{The FCN-8 model is presented in block diagram that consist of 5 fully convolutional network blocks, and 2 up-sampling blocks. The model receives input as a SAR image and perform pixel-wise classification to output a labeled image. }\label{fig:fcn8}}
\end{figure}
\subsection{Oil Spill Detection  (abbreviated as Oil)}
Detecting the oil spill is of paramount importance to have a safe and clean O\&G Industry 4.0. The accuracy of DNN-based oil spill detection systems has been promising \cite{krestenitis2019oil}. We utilize a detection system that operates based on the FCN-8 model \cite{long2015fully}, which is depicted in Figure~\ref{fig:fcn8}. As we can see, the model contains five Fully Convolutional Network (FCN) blocks and two up-sampling blocks that collectively perform semantic segmentation (\ie classifying every pixel) of an input image and output a labeled image. The FCN-8 model functions based on the satellite (a.k.a. SAR) \cite{huang2020classification} images. We configure the analysis to obtain the inference time of 110 SAR images collected by MKLab \cite{krestenitis2019oil}.

\subsection{Acoustic Impedance Estimation  (abbreviated as AIE)}
\begin{figure}[h!]
	\centering	
	\includegraphics[scale=0.46]{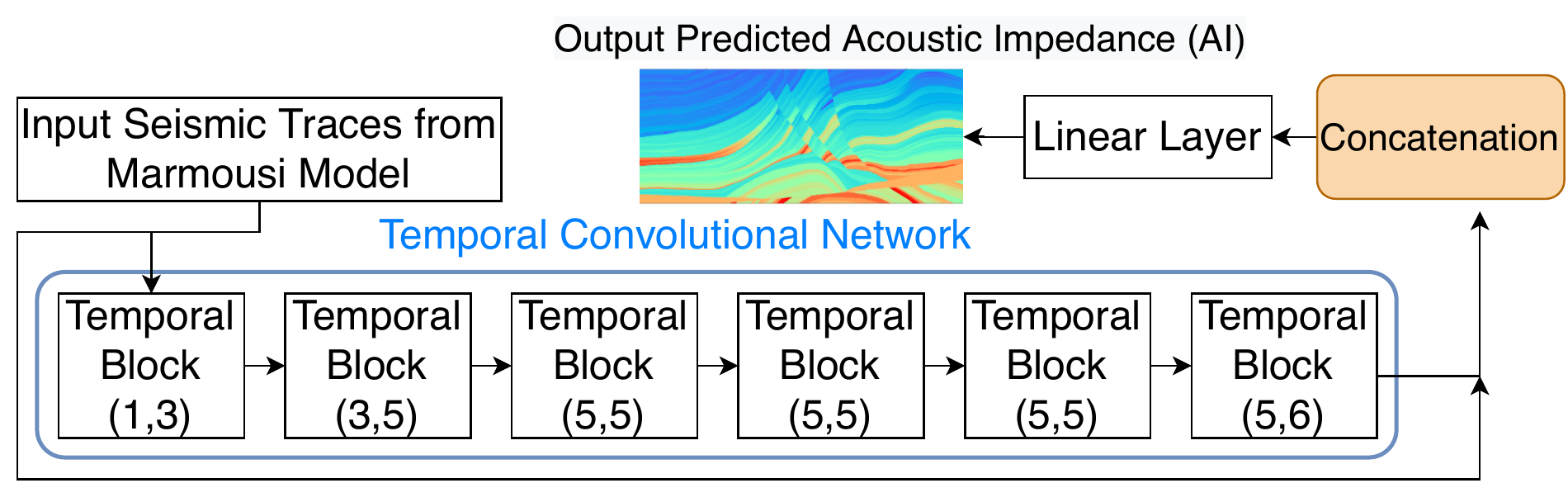}
	\caption{\small{Schematic view of Temporal Convolutional Network (TCN) model that consists of six temporal blocks, the input data, and the output in form of the predicted AI.}}\label{fig:tcn}
\end{figure}

Estimating acoustic impedance (AI) from seismic data is an important step in O\&G exploration. To estimate AI from seismic data, we utilize a solution functions based on the temporal convolutional network \cite{mustafa2019estimation}, shown in Figure \ref{fig:tcn}. The solution outperforms others in terms of gradient vanishing and overfitting. Marmousi 2 dataset \cite{marmousi} is used to estimate AI. 

\section{Cloud Computing Platforms for Industry 4.0}\label{insType}
\paragraph{Amazon Cloud}
AWS is a pioneer in the Cloud computing industry and provides more than 175 services, including Amazon EC2 \cite{varia2014overview}, across a large set of distributed data centers. Amazon EC2 provides inconsistently heterogeneous machines (\eg CPU, GPU, and Inferentia) in form of various VM instance types (\eg general purpose, compute-optimized, and machine learning (ML)). Within each VM type, a range of VM configurations (\eg \texttt{large, xlarge, 2xlarge}) are offered that reflect the consistent heterogeneity within that VM type. To realize the impact of machine heterogeneity on the inference time of various applications, we choose one representative VM type of each offered machine type. Table~\ref{awsInstances} represents the type of machines and their specification in terms of number of cores and memory. We note that all the machine types use SSD storage units. Although General Purpose machines are not considered suitable for latency-sensitive DNN-based applications, the reason we study them is their similarity to the specifications of machine types often used in the edge computing platforms. As such, considering these types of machines (and similarly $m1.small$ in the Chameleon cloud) makes the results of this study applicable to cases where edge computing is employed for latency-sensitive applications \cite{icfec19vaughan}.

\begin{table}[h!]
\centering
\scalebox{0.85}
{\begin{tabular}{|l|l|c|c|c|c|c|c|}
\hline
\textbf{Machine Type}           & \textbf{VM Config.} & \multicolumn{1}{l|}{\textbf{vCPU}} & \multicolumn{1}{l|}{\textbf{GPU}} & \multicolumn{1}{r|}{\textbf{Mem. (GB)}}\\ \hline
Mem. Optimized          & \texttt{r5d.xlarge}     & 4                                  & 0                                 & 32                                                                                                                        \\ \hline
ML Optimized & \texttt{inf1.xlarge}             & 4                                  & 0                                 & 8                                                                                                                                                             \\ \hline
GPU                      & \texttt{g4dn.xlarge}             & 4                                  & 1                                 & 16                                                                                                                                         \\ \hline
General Purpose            & \texttt{m5ad.xlarge}             & 4                                  & 0                                 & 16                                                                                                                                                      \\ \hline
Comp. Optimized          & \texttt{c5d.xlarge}              & 4                                  & 0                                 & 8                                                                                                                                        \\ \hline
\end{tabular}}
\caption{Heterogeneous machine types and VM configurations in Amazon EC2 that are considered for performance modeling of DNN-based applications. In this table, ML Optimized represents Inferentia machine type offered by AWS.}
\label{awsInstances}
\end{table}

\begin{table}[h!]
\centering
{\begin{tabular}{l||c|c}
\textbf{VM Config.}       & \textbf{vCPUS} & \textbf{Mem. (GB)}   \\ \hline\hline
\texttt{m1.xlarge} & 8 & 16 \\ \hline
\texttt{m1.large}  & 4  & 8  \\ \hline
\texttt{m1.medium}   & 2  & 4 \\ \hline
\texttt{m1.small}  & 1  & 2  \\ \hline
\end{tabular}}
\caption{Various VM flavors in Chameleon cloud are configured to represent a consistently heterogeneous system.}
\label{tab:chameleon}
\end{table}

\begin{figure*}[h!]
\centering
\includegraphics[width=1\textwidth]{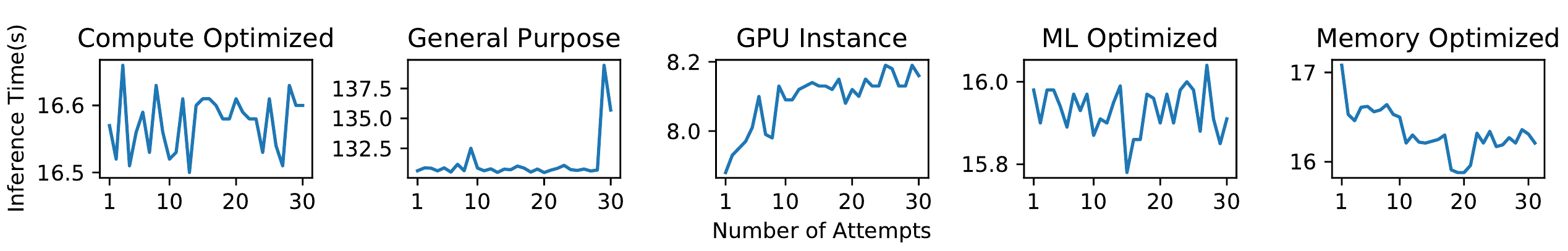}
\caption{\small{The stochastic nature of inference execution time of oil spill application while running on heterogeneous VMs in the AWS. For every VM instance, the oil spill detection application is executed 30 times and those executions are plotted as number of attempts along x-axis. The y-axis represents the inference time for every attempts.}}
\label{fig:stochasticoilaws}
\centering
\end{figure*}


\paragraph{Chameleon Cloud}
Chameleon cloud \cite{chameleon} is a large-scale public cloud maintained by National Science Foundation (NSF). Chameleon cloud supports VM-based heterogeneous computing. It offers the flexibility to manage the compute, memory, and storage capacity of the VM instances. In this study, we use the Chameleon cloud to configure a set of consistently heterogeneous machines. We configure four VM flavors, namely \texttt{m1.xlarge, m1.large, m1.medium}, and \texttt{m1.small}, as detailed in Table~\ref{tab:chameleon}.
We note that VMs offered by Chameleon cloud uses HDD unit as storage.

\section{Environmental Setup for Performance Modeling}\label{expSetup}
The focus of this study is on latency-sensitive DNN-based applications that are widely used in Industry 4.0. Accordingly, we consider a dynamic (online) system that is already loaded with pre-trained DNN-based applications, explained in the previous section, and executes arriving requests on the pertinent application. This means that we measure the hot start inference time \cite{ogden2019characterizing} in the considered applications. The DNN-based applications mostly use TensorFlow, and the VMs both in AWS and Chameleon are configured to use the framework on top of Ubuntu 18.04.

Our initial evaluations in AWS (shown in Figure \ref{fig:stochasticoilaws}) demonstrate that, in different attempts, the inference execution time of an application (Oil Spill) on the same machine type can be highly stochastic. Similar stochasticity is found for chameleon cloud while we run the oil spill detection application 30 times within same VM instance. Hence to capture this randomness (aka consistent heterogeneity) that is caused by several factors, such as transient failures or multi-tenancy \cite{moradi2019adaptive,performanceXiambo}, we base our analysis on 30 times execution of the same request within same VM.\looseness=-1

\begin{table}[h!]
\centering
\scalebox{0.63}
{\begin{tabular}{|c|c|c|c|c|c|}
\hline
\multicolumn{6}{|c|}{\textbf{Execution Time Distribution with Shapiro-Wilk Test in AWS Cloud}} \\ \hline
\textbf{App. Type} & \textbf{\texttt{Mem. Opt.}} & \textbf{\texttt{ML Opt.}} & \textbf{\texttt{GPU}} & \textbf{\texttt{Gen. Pur.}} & \textbf{\texttt{Compt. Opt.}} \\ \hline
\textit{Fire} & \begin{tabular}[c]{@{}c@{}}Not Gaussian\\ (P=2.73$e^{-16}$)\end{tabular} & \begin{tabular}[c]{@{}c@{}}Not Gaussian\\ (P=5.42$e^{-16}$)\end{tabular} & \begin{tabular}[c]{@{}c@{}}Not Gaussian\\ (P=6.59$e^{-16}$)\end{tabular} & \begin{tabular}[c]{@{}c@{}}Not Gaussian\\ (P=2.06$e^{-13}$)\end{tabular} & \begin{tabular}[c]{@{}c@{}}Not Gaussian\\ (P=3.9$e^{-16}$)\end{tabular} \\ \hline
\textit{\begin{tabular}[c]{@{}c@{}}HAR\end{tabular}} & \begin{tabular}[c]{@{}c@{}}Not Gaussian\\ (P=7.12$e^{-8}$)\end{tabular} & \begin{tabular}[c]{@{}c@{}}Not Gaussian\\ (P=1.04$e^{-8}$)\end{tabular} & \begin{tabular}[c]{@{}c@{}}Gaussian\\ (P=0.19)\end{tabular} & \begin{tabular}[c]{@{}c@{}}Not Gaussian\\ (P=1.76$e^{-8}$)\end{tabular} & \begin{tabular}[c]{@{}c@{}}Not Gaussian\\ (P=0.4.62$e^{-5}$)\end{tabular} \\ \hline
\textit{\begin{tabular}[c]{@{}c@{}}Oil\end{tabular}} & \begin{tabular}[c]{@{}c@{}}Not Gaussian\\ (P=8$e^{-4}$)\end{tabular} & \begin{tabular}[c]{@{}c@{}}Not Gaussian\\ (P=2.9$e^{-16}$)\end{tabular} & \begin{tabular}[c]{@{}c@{}}Not Gaussian\\ (P=0.012)\end{tabular} & \begin{tabular}[c]{@{}c@{}}Not Gaussian\\ (P=1.27$e^{-16}$)\end{tabular} & \begin{tabular}[c]{@{}c@{}}Not Gaussian\\ (P=5.86$e^{-14}$)\end{tabular} \\ \hline
\textit{\begin{tabular}[c]{@{}c@{}}AIE\end{tabular}} & \begin{tabular}[c]{@{}c@{}}Gaussian\\ (P=0.46)\end{tabular} & \begin{tabular}[c]{@{}c@{}}Gaussian\\ (P=0.23)\end{tabular} & \begin{tabular}[c]{@{}c@{}}Gaussian\\ (P=0.08)\end{tabular} & \begin{tabular}[c]{@{}c@{}}Not Gaussian\\ (P=1.99$e^{-10}$)\end{tabular} & \begin{tabular}[c]{@{}c@{}}Gaussian\\ (P=0.96)\end{tabular} \\ \hline
\end{tabular}}
\caption{\small{The execution time distributions of DNN-based applications in AWS clouds machines using Shapiro-Wilk test.}}
\label{tab:awsShaprio}
\end{table}

\begin{table}[h!]
\centering
\scalebox{0.78} 
{\begin{tabular}{|c|c|c|c|c|}
\hline
\multicolumn{5}{|c|}{\textbf{Execution Time Distribution with Shapiro-Wilk Test in Chemeleon}} \\ \hline
 \textbf{App. Type} & \textbf{\texttt{m1.xlarge}} & \textbf{\texttt{m1.large}} & \textbf{\texttt{m1.medium}} & \textbf{\texttt{m1.small}} \\ \hline
\textit{Fire} & \begin{tabular}[c]{@{}c@{}}Not Gaussian\\(P=4.05$e^{-5}$) \end{tabular}&\begin{tabular}[c]{@{}c@{}} Not Gaussian\\(P=1.$e^{-4}$) \end{tabular}&\begin{tabular}[c]{@{}c@{}} Not Gaussian\\(P=7.58$e^{-6}$) \end{tabular}&\begin{tabular}[c]{@{}c@{}} Not Gaussian\\(P=1.32$e^{-7}$) \end{tabular}\\ \hline
\textit{\begin{tabular}[c]{@{}c@{}}HAR\end{tabular}} &\begin{tabular}[c]{@{}c@{}} Gaussian \\ (P=0.74) \end{tabular}&\begin{tabular}[c]{@{}c@{}} Not Gaussian \\ (P=0.02) \end{tabular}& \begin{tabular}[c]{@{}c@{}}Gaussian \\ (P=0.18) \end{tabular}&\begin{tabular}[c]{@{}c@{}} Gaussian \\ (P=0.84) \end{tabular}\\ \hline
\textit{\begin{tabular}[c]{@{}c@{}}Oil\end{tabular}} &\begin{tabular}[c]{@{}c@{}} Not Gaussian \\ (P=0.01) \end{tabular}&\begin{tabular}[c]{@{}c@{}} Not Gaussian \\ (P=5.5$e^{-7}$) \end{tabular}& \begin{tabular}[c]{@{}c@{}}Not Gaussian \\ (P=0.01)\end{tabular}& N/A \\ \hline
\textit{\begin{tabular}[c]{@{}c@{}}AIE\end{tabular}} & \begin{tabular}[c]{@{}c@{}}Not Gaussian \\ (P=2.77$e^{-10}$)\end{tabular}& \begin{tabular}[c]{@{}c@{}}Not Gaussian \\ (P= 3.46$e^{-6}$)\end{tabular}& \begin{tabular}[c]{@{}c@{}} Not Gaussian \\ (P= 1.4$e^{-4}$)\end{tabular}& \begin{tabular}[c]{@{}c@{}}Not Gaussian \\ (P=2.46$e^{-6}$)\end{tabular}\\ \hline
\end{tabular}}
\caption{\small{The execution time distributions of DNN applications in Chameleon cloud using Shapiro-Wilk test.}}
\label{tab:chamShap}
\end{table}

\section{Application-Centric Analysis of Inference Time}\label{result}
\subsection{Overview}
In this part, we capture the inference time of the four DNN applications and try to identify their underlying statistical distributions using various statistical methods. Then, to describe the behavior of inference execution time using a single metric, we explore the central tendency of the distributions. 

\subsection{Statistical Distribution of Inference Execution Time}
Among various statistical methods, the normality tests are widely employed to understand the distribution of the collected samples. Hence, we first use Shapiro-Wilk test \cite{hanusz2016shapiro} to verify the normality of the inference time distribution on each machine type. Next, we employ Kolmogorov-Smirnov test \cite{chakravarty1967handbook} to find the best fit distribution based on the sampled inference execution times.

\subsubsection{Shapiro-Wilk Test to Verify Normality of the Sampled Data}
The null hypothesis is that the inference execution times are normally distributed.
~To understand whether a random sample comes from a normal distribution, we perform the Shapiro-Wilk test. The result of this test is considered as $W$, whose low value (lower than $w_\alpha$ threshold) indicates that the sampled data are not normally distributed and vice versa.
The value of $W$ is used to perform the significant testing (\ie calculating P-value). The higher P-value, especially greater than a threshold value (typically 0.05), supports the null hypothesis that the sampled data are normally distributed. 

The results of Shapiro-Wilk test on the collected inference times for AWS are presented in Table \ref{tab:awsShaprio}, where columns present the various machine types and rows define the application types. The table reflects that our initial assumption is not totally valid. The Shapiro-Wilk test result for the Chameleon cloud, depicted in Table \ref{tab:chamShap}, shows that for only three of the total cases, the normality assumption holds. Considering the lack of normality in several cases, in the next section, we utilize Kolmogorov-Smirnov test to more granularly explore the best fitting distribution for the inference time of each application and also cross validate the prior results we obtained using another statistical method.
\begin{table}[h!]
\centering
\scalebox{0.67}
{\begin{tabular}{|c|c|c|c|c|c|}
\hline
\multicolumn{6}{|c|}{\textbf{Execution Time Distribution with Kolmogorov-Smirnov Test in AWS Cloud}} \\ \hline
 \textbf{App. Type} & \textbf{\texttt{Mem. Opt.}} & \textbf{\texttt{ML Opt.}} & \textbf{\texttt{GPU}} & \textbf{\texttt{Gen. Pur.}} & \textbf{\texttt{Compt. Opt.}} \\ \hline
\textit{Fire} & No Distr. & No Distr. & No Distr. & No Distr. & No Distr. \\ \hline
\textit{\begin{tabular}[c]{@{}c@{}}HAR\end{tabular}} & \begin{tabular}[c]{@{}c@{}}Student's t\\ (P=0.08)\end{tabular} & \begin{tabular}[c]{@{}c@{}}Student's t\\ (P=0.77)\end{tabular} & \begin{tabular}[c]{@{}c@{}}Student's t\\ (P=0.99)\end{tabular} & \begin{tabular}[c]{@{}c@{}}Student's t\\ (P=0.57)\end{tabular} & \begin{tabular}[c]{@{}c@{}}Student's t\\ (P=0.95)\end{tabular} \\ \hline
\textit{\begin{tabular}[c]{@{}c@{}}Oil\end{tabular}} & \begin{tabular}[c]{@{}c@{}}Student's t\\ (P=0.44)\end{tabular} & \begin{tabular}[c]{@{}c@{}}Student's t\\ (P=0.96)\end{tabular} & \begin{tabular}[c]{@{}c@{}}Student's t\\ (P=0.5)\end{tabular} & \begin{tabular}[c]{@{}c@{}}Student's t\\ (P=0.20)\end{tabular} & \begin{tabular}[c]{@{}c@{}}Exponential\\ (P=0.21)\end{tabular} \\ \hline
\textit{\begin{tabular}[c]{@{}c@{}}AIE\end{tabular}} & \begin{tabular}[c]{@{}c@{}}Normal\\ (P=0.99)\end{tabular} & \begin{tabular}[c]{@{}c@{}}Normal\\ (P=0.54)\end{tabular} & \begin{tabular}[c]{@{}c@{}}Normal\\ (P=0.47)\end{tabular} & \begin{tabular}[c]{@{}c@{}}Exponential\\ (P=0.16)\end{tabular} & \begin{tabular}[c]{@{}c@{}}Normal\\ (P=0.99)\end{tabular} \\ \hline
\end{tabular}}
\caption{\small{Inference time distributions of DNN-based applications in AWS cloud machines using Kolmogorov-Smirnov test.}} 
\label{tab:kolmoAWS}
\end{table}
\subsubsection{Kolmogorov-Smirnov Goodness of Fit Test}
The null hypothesis for the Kolmogorov-Smirnov test is that the inference times of a certain application type on a given machine type follows a statistical distribution. The Kolmogorov-Smirnov Goodness of Fit test (a.k.a. \emph{K-S test}) identifies whether a set of samples derived from a population fits to a specific distribution. Precisely, the test measures the largest vertical distance (called test statistic $D$) between a known hypothetical probability distribution and the distribution generated by the observed inference times (a.k.a. empirical distribution function (EDF)). Then, if $D$ is greater than the critical value obtained from the K-S test P-Value table, then the null hypothesis is rejected.

The results of the K-S test on the observed inference times in AWS and Chameleon clouds are depicted in Table \ref{tab:kolmoAWS} and \ref{tab:chamKol}, respectively. From Table \ref{tab:kolmoAWS}, we find that, in AWS, majority of the entries either represent Normal distribution or Student's t-distribution that exposes similar properties. However, we observe that the inference time of Fire Detection application does not follow any particular distribution with an acceptable P-Value. We also observe that the inference times of both Oil Spill application on Compute Optimized machine and AIE application on General Purpose machine follow Exponential distribution. However, low P-Value in both of these cases indicate a weak acceptance of the null hypothesis.
\vspace{-4mm}
\begin{table}[h!]
\centering
\scalebox{0.75}
{\begin{tabular}{|l|l|l|l|l|}
\hline
\multicolumn{5}{|c|}{\textbf{Execution Time Distribution with Kolmogorov-Smirnov test in Chameleon}} \\ \hline
\textbf{App. Type}  & \textbf{\texttt{m1.xlarge}}   & \textbf{\texttt{m1.large}}  & \textbf{\texttt{m1.medium}} & \textbf{\texttt{m1.small}} \\ \hline
\textit{Fire}                   & No Distr   & No Distr & No Distr & Log-normal  \\ \hline
\textit{HAR}       & \begin{tabular}[c]{@{}c@{}}Normal \\ (P=0.98)\end{tabular}  & \begin{tabular}[c]{@{}c@{}}Student's t\\ (P=0.88) \end{tabular}&\begin{tabular}[c]{@{}c@{}} Normal \\ (P=0.66)\end{tabular}& \begin{tabular}[c]{@{}c@{}}Normal \\ (P=0.96)   \end{tabular}  \\ \hline
\textit{Oil}              & \begin{tabular}[c]{@{}c@{}}Log-normal\\(P=0.36) \end{tabular} & \begin{tabular}[c]{@{}c@{}}Log-normal\\(P=0.99) \end{tabular}& \begin{tabular}[c]{@{}c@{}}Log-normal\\(P=0.81)\end{tabular}      & N/A            \\ \hline
\textit{AIE}                & \begin{tabular}[c]{@{}c@{}} Student's t \\(P= 0.47)  \end{tabular} &\begin{tabular}[c]{@{}c@{}} Student's t \\(P=0.12) \end{tabular}   & \begin{tabular}[c]{@{}c@{}}Student's t\\(P=0.25) \end{tabular}  & \begin{tabular}[c]{@{}c@{}}Student's t\\(P=0.83) \end{tabular} \\ \hline
\end{tabular}}
\caption{\small{Inference time distributions of DNN-based applications in Chameleon's machines using the K-S test.}}
\label{tab:chamKol}
\end{table}

On the contrary, Table \ref{tab:chamKol} reflects the dominance of Log-normal (\ie the logarithm of the random variable is normally distributed) and Student's t-distribution over other distributions in the Chameleon cloud. Analysing the execution traces shows us that the inference times in Chameleon are predominantly larger than the ones in AWS that causes right-skewed property, hence, the distribution tends to Log-normal. Consistent to AWS observations, we see that Fire Detection application, in most of the cases, does not follow any distribution. Our further analysis showed that the lack of distribution is because of variety (\eg frame rate and resolution) in the input videos. In fact, when we reduced the degree of freedom in the input videos limited them to those with the same configuration (frame-rate), we noticed the inference time follows a Log-normal distribution. The observation shows that the characteristics and variation of input data can be decisive on the statistical behavior of inference times (mentioned in highlighted insight). 
We note that Oil Spill application cannot be run on \texttt{m1.small} machine owing to its limited memory.
\begin{table}[h!]
\centering
\scalebox{0.68}
{\begin{tabular}{|c|c|c|c|c|c|}
\hline
\multicolumn{6}{|c|}{\textbf{Mean and Standarad Deviation of Inference Execution Times (ms) in AWS}} \\ \hline
\textbf{App. Type} & \textbf{\texttt{Mem. Opt.}} & \textbf{\texttt{ML Opt.}} & \textbf{\texttt{GPU}} & \textbf{\texttt{Gen. Pur.}} & \textbf{\texttt{Compt. Opt.}} \\ \hline
\textit{Fire} & \begin{tabular}[c]{@{}c@{}}$\mu$=1461.8\\ $\sigma$=457.3\end{tabular} & \begin{tabular}[c]{@{}c@{}}$\mu$=1281.7\\ $\sigma$=387.93\end{tabular} & \begin{tabular}[c]{@{}c@{}}$\mu$=1349.5\\ $\sigma$=418.9\end{tabular} & \begin{tabular}[c]{@{}c@{}}$\mu$ =1534.8\\ $\sigma$=494.7\end{tabular} & \begin{tabular}[c]{@{}c@{}}$\mu$=1421.4\\ $\sigma$=441.8\end{tabular} \\ \hline
\textit{\begin{tabular}[c]{@{}c@{}}HAR\end{tabular}} & \begin{tabular}[c]{@{}c@{}}$\mu$=1.27\\ $\sigma$=0.082\end{tabular} & \begin{tabular}[c]{@{}c@{}}$\mu$=0.66\\ $\sigma$=0.006\end{tabular} & \begin{tabular}[c]{@{}c@{}}$\mu$=0.51\\ $\sigma$=0.006\end{tabular} & \begin{tabular}[c]{@{}c@{}}$\mu$ =1.17\\ $\sigma$=0.042\end{tabular} & \begin{tabular}[c]{@{}c@{}}$\mu$=0.66\\ $\sigma$=0.003\end{tabular} \\ \hline
\textit{\begin{tabular}[c]{@{}c@{}}Oil\end{tabular}} & \begin{tabular}[c]{@{}c@{}}$\mu$=269.9\\ $\sigma$=1.01\end{tabular} & \begin{tabular}[c]{@{}c@{}}$\mu$=218.8\\ $\sigma$=0.66\end{tabular} & \begin{tabular}[c]{@{}c@{}}$\mu$=65.98\\ $\sigma$=0.47\end{tabular} & \begin{tabular}[c]{@{}c@{}}$\mu$=667.1\\ $\sigma$=2.26\end{tabular} & \begin{tabular}[c]{@{}c@{}}$\mu$=242.9\\ $\sigma$=0.68\end{tabular} \\ \hline
\textit{\begin{tabular}[c]{@{}c@{}}AIE\end{tabular}} & \begin{tabular}[c]{@{}c@{}}$\mu$=7.02\\ $\sigma$=0.02\end{tabular} & \begin{tabular}[c]{@{}c@{}}$\mu$=6.41\\ $\sigma$=0.03\end{tabular} & \begin{tabular}[c]{@{}c@{}}$\mu$=7.55\\ $\sigma$=0.04\end{tabular} & \begin{tabular}[c]{@{}c@{}}$\mu$=9.35\\ $\sigma$=0.06\end{tabular} & \begin{tabular}[c]{@{}c@{}}$\mu$=7.95\\ $\sigma$=0.02\end{tabular} \\ \hline
\end{tabular}}
\caption{\small{The measurement of central tendency metric ($\mu$), and data dispersion metric ($\sigma$) on the observed inference times in AWS.}}
\label{tab:meanAWS}
\end{table}

\noindent\textbf{Insights:}
The key insights of the analysis we conducted on identifying the distribution of inference time are as follows:
\begin{mdframed}[backgroundcolor=blue!20] \begin{itemize}
    \item Shapiro-Wilk test for AWS and Chameleon rejects the null hypothesis that the inference times of DNN-based applications follow the Normal distribution.
    \item The K-S test reflects the dominance of Student's t-distribution of inference time in both AWS (Table \ref{tab:kolmoAWS}), and Chameleon (Table \ref{tab:chamKol}). 
    \item Various configurations of input data is decisive on the statistical distribution of the inference time. 
\end{itemize}
\end{mdframed}
\vspace{-2mm}
\subsection{Analysis of Central Tendency and Dispersion Measures}
Leveraging the statistical distributions of inference times, in this part, we explore their central tendency metric that summarizes the behavior of collected observations in a single value. In addition, to analyze the statistical dispersion of the observations, we measure the standard deviation of the inference times. In particular, we estimate the arithmetic mean and standard deviation of the inference times. The central tendency metric of inference times for AWS and Chameleon clouds are shown in Tables~\ref{tab:meanAWS} and~\ref{tab:meanCham}, respectively. The \textbf{key insights} are as follows:
\begin{mdframed}[backgroundcolor=blue!20] 
\begin{itemize}
    \item Machine Learning Optimized and GPU instances often outperform other AWS machine types.
    \item In both clouds, the inference times of Fire and Oil experience a higher standard deviation in compare with other applications. The high uncertainty is attributed to the characteristics of their input data; Oil Spill input images suffer from class imbalance \cite{krestenitis2019oil}, whereas, Fire input videos are highly variant. 
    \item In Chameleon VMs with a consistent heterogeneity, DNN applications with dense network models (\eg Oil and Fire) can take advantage of powerful machines (\eg \texttt{m1.xlarge}) to significantly reduce their inference times.
    \item Overall, AWS offers a lower inference time than Chameleon. The reason is utilizing SSD units in AWS rather than HDD in Chameleon. In addition, we noticed that Chameleon experiences more transient failures that can slow down the applications.
\end{itemize}
\end{mdframed}
\vspace{-4mm}
\begin{table}[h!]
\centering
\scalebox{0.82}
{\begin{tabular}{|l|l|l|l|l|}
\hline
\multicolumn{5}{|l|}{\textbf{Mean and Std. of Inference Execution Times (ms) in Chameleon}} \\ \hline
\textbf{App. Type} & \textbf{\texttt{m1.xlarge}} & \textbf{\texttt{m1.large}} & \textbf{\texttt{m1.medium}} & \textbf{\texttt{m1.small}} \\ \hline
\textit{Fire} & \begin{tabular}[c]{@{}l@{}}$\mu$=2155.20\\ $\sigma$=725.48\end{tabular} & \begin{tabular}[c]{@{}l@{}}$\mu$=2213.30\\ $\sigma$=731.50\end{tabular} & \begin{tabular}[c]{@{}l@{}}$\mu$=2330.80\\ $\sigma$=742.20\end{tabular} & \begin{tabular}[c]{@{}l@{}}$\mu$=3184.80\\ $\sigma$=1033.30\end{tabular} \\ \hline
\textit{\begin{tabular}[c]{@{}l@{}}HAR\end{tabular}} & \begin{tabular}[c]{@{}l@{}}$\mu$=22.14\\ $\sigma$=0.76\end{tabular} & \begin{tabular}[c]{@{}l@{}}$\mu$=47.69\\ $\sigma$=1.26\end{tabular} & \begin{tabular}[c]{@{}l@{}}$\mu$=49.24\\ $\sigma$=0.57\end{tabular} & \begin{tabular}[c]{@{}l@{}}$\mu$=52.69\\ $\sigma$=0.78\end{tabular} \\ \hline
\textit{\begin{tabular}[c]{@{}l@{}}Oil\end{tabular}} & \begin{tabular}[c]{@{}l@{}}$\mu$=147.16\\ $\sigma$=5.23\end{tabular} & \begin{tabular}[c]{@{}l@{}}$\mu$=222.22\\ $\sigma$=2.89\end{tabular} & \begin{tabular}[c]{@{}l@{}}$\mu$=412.78\\ $\sigma$=4.99\end{tabular} & N/A \\ \hline
\textit{\begin{tabular}[c]{@{}l@{}}AIE\end{tabular}} & \begin{tabular}[c]{@{}l@{}}$\mu$=6.23\\ $\sigma$=0.25\end{tabular} & \begin{tabular}[c]{@{}l@{}}$\mu$=6.23\\ $\sigma$=0.15\end{tabular} & \begin{tabular}[c]{@{}l@{}}$\mu$=6.40\\ $\sigma$=0.13\end{tabular} & \begin{tabular}[c]{@{}l@{}}$\mu$=7.72\\ $\sigma$=0.24\end{tabular} \\ \hline
\end{tabular}}
\caption{\small{Central tendency metric ($\mu$), and data dispersion metric ($\sigma$) of the  inference times in the Chameleon cloud.}}
\label{tab:meanCham}
\end{table}

\begin{table}[h!]
\centering
\scalebox{0.65}
{\begin{tabular}{|c|c|c|c|c|c|}
\hline
\multicolumn{6}{|c|}{\textbf{The MIPS for DNN Applications in AWS Cloud}} \\ \hline
\textbf{App. Type} & \textbf{\texttt{Mem. Opt.}} & \textbf{\texttt{ML Opt.}} & \textbf{\texttt{GPU}} & \textbf{\texttt{Gen. Pur.}} & \textbf{\texttt{Compt. Opt.}} \\ \hline
\textit{Fire} & 1938.63 & 2196.35 & 2092.72 & 1862.04 & 1989.56 \\ \hline
\textit{\begin{tabular}[c]{@{}c@{}}HAR\end{tabular}} & 838640.65 & 1595874.34 & 2040057.33 & 891754.48 & 1581709.12 \\ \hline
\textit{\begin{tabular}[c]{@{}c@{}}Oil\end{tabular}} & 164.54 & 168.58 & 331.98 & 20.46 & 162.01 \\ \hline
\textit{\begin{tabular}[c]{@{}c@{}}AIE\end{tabular}} & 145.58 & 180.28 & 150.25 & 131.25 & 160.32 \\ \hline
\end{tabular}}
\caption{\small{MIPS values of heterogeneous machines in AWS for each DNN-based application.}}
\label{mipsAWS}
\end{table}

\begin{table}[h!]
\centering
\scalebox{0.80}
{\begin{tabular}{|l|l|l|l|l|}
\hline
\multicolumn{5}{|c|}{\textbf{The MIPS for DNN Applications in Chameleon}}                   \\ \hline
\textbf{App. Types} & \textbf{\texttt{m1.xlarge}} & \textbf{\texttt{m1.large}} & \textbf{\texttt{m1.medium}} & \textbf{\texttt{m1.small}} \\ \hline
\textit{Fire}                                   & 1327.81 & 1282.33  & 1249.63 & 871.36\\ \hline
\textit{HAR}                       & 91.78    & 102.51   & 124.76 & 136.62  \\ \hline
\textit{Oil}                              & 18267.35   & 11233.41  & 6243.94  & N/A    \\ \hline
\textit{AIE}                                & 246366.52   & 249551.29 & 236300.93 & 201807.49 \\ \hline
\end{tabular}}
\caption{\small{MIPS vales for heterogeneous machines on Chameleon cloud for each DNN-based application.}}
\label{mipsCham}
\end{table}

\section{Resource-Centric Analysis of Inference Time}\label{resCentric}
In performance analysis of computing systems, a rate-based metric \cite{lilja2005measuring} is defined as the normalization of number of computer instructions executed to a standard time unit. MIPS is a popular rate-based metric that allows comparison of computing speed across two or more computing systems. Given that computing systems (\eg AWS ML Optimized and GPU) increasingly use instruction-level facilities for ML applications, our objective in this part is to analyze the performance of different machine types in processing DNN-based applications. The results of this analysis can be of particular interest to researchers and cloud solution architects whose endeavor is to develop tailored resource allocation solutions for Industry 4.0 use cases.  As for rate-based metrics we do not assume any distribution \cite{patil2010using}, we conduct a non-parametric approach. In addition to MIPS, we provide the range of MIPS in form of \textit{Confidence Intervals} (CI) for each case. \looseness=-1

Let application $i$ with $n_i$ instructions have $t_{im}$ inference time on machine $m$. Then, MIPS of machine $m$ to execute the application is defined as $MIPS_{mi} = n_i/(t_{im}\times 10^6)$. Hence, before calculating MIPS for any machine, we need to estimate the number of instructions ($n$) of each DNN-based application. For that purpose, we execute each task $t_i$ on a machine whose MIPS is known and estimated $n_i$. Then, for each machine $m$, we measure $t_{im}$ and subsequently calculate $MIPS_{mi}$. Tables~\ref{mipsAWS} and~\ref{mipsCham} show the MIPS values for AWS and Chameleon, respectively.\looseness=-1  

To measure the confidence intervals (CI) of MIPS for each application type in each machine type, we use the non-parametric statistical methods \cite{patil2010using} that perform prediction based on the sample data without making any assumption about their underlying distributions. As we deal with a rate-based metric, we use harmonic mean that offers a precise analysis for this type of metric rather than the arithmetic mean. We utilize Jackknife \cite{patil2010using} re-sampling method and validate it using Bootstrap \cite{patil2010using}, which is another well-known re-sampling method. Both of these methods employ harmonic mean to measure the confidence intervals of MIPS. 

\begin{table}[h!]
\centering
\scalebox{0.63}
{\begin{tabular}{|l|l|l|l|l|l|}
\hline
\multicolumn{6}{|c|}{\textbf{CI of MIPS using Jackknife Method in AWS cloud}}                                                                                                                                                                                                                                                                                     \\ \hline
\textbf{App. Type} & \textbf{\texttt{Mem. Opt.}}                                                    & \textbf{\texttt{ML Opt.}}                                          & \textbf{\texttt{GPU}}                                                                 & \textbf{\texttt{Gen. Pur.}}                                                     & \textbf{\texttt{Compt. Opt.}}                                                   \\ \hline
\textit{Fire} & \begin{tabular}[c]{@{}l@{}}[1549.42,\\ 1975.65]\end{tabular} & \begin{tabular}[c]{@{}l@{}}[1770.81,\\2243.04]\end{tabular} & \begin{tabular}[c]{@{}l@{}}[1671.78,\\2131.66]\end{tabular} & \begin{tabular}[c]{@{}l@{}}[1465.31,\\1889.77]\end{tabular} & \begin{tabular}[c]{@{}l@{}}[1594.78,\\2028.36]\end{tabular} \\ \hline
\textit{HAR}      & \begin{tabular}[c]{@{}l@{}}[812040.26,\\ 856355.96]\end{tabular}   & \begin{tabular}[c]{@{}l@{}}[1592214.75,\\ 1599426.64]\end{tabular}   & \begin{tabular}[c]{@{}l@{}}[2033084.47,\\ 2046727.57]\end{tabular} & \begin{tabular}[c]{@{}l@{}}[880417.69,\\ 901345.49]\end{tabular} & \begin{tabular}[c]{@{}l@{}}[1580275.10,\\ 1585598.85]\end{tabular}   \\ \hline
\textit{Oil}             & \begin{tabular}[c]{@{}l@{}}[163.55,\\165.47]\end{tabular} & \begin{tabular}[c]{@{}l@{}}[168.36,\\168.81]\end{tabular} & \begin{tabular}[c]{@{}l@{}}[330.68,\\333.22]\end{tabular} & \begin{tabular}[c]{@{}l@{}}[20.35,\\20.57]\end{tabular}   & \begin{tabular}[c]{@{}l@{}}[161.86,\\162.17]\end{tabular} \\ \hline
\textit{AIE}               & \begin{tabular}[c]{@{}l@{}}[139.02,\\141.04]\end{tabular} & \begin{tabular}[c]{@{}l@{}}[155.56,\\156.01]\end{tabular} & \begin{tabular}[c]{@{}l@{}}[141.57,\\142.03]\end{tabular} & \begin{tabular}[c]{@{}l@{}}[118.06,\\119.82]\end{tabular} & \begin{tabular}[c]{@{}l@{}}[148.35,\\149.00]\end{tabular} \\ \hline
\end{tabular}}
\caption{\small{The confidence intervals of MIPS values for DNN-based applications in AWS machines, resulted from Jackknife re-sampling method.}}
\label{jackAWS}
\end{table}

\begin{table}[h!]
\centering
\scalebox{0.80}
{\begin{tabular}{|l|l|l|l|l|}
    \hline
    \multicolumn{5}{|c|}{\textbf{CI of MIPS using Jackknife Method in Chameleon Cloud}}                                                                                                                                                                                                                                \\ \hline
    \textbf{App. Type} & \textbf{\texttt{m1.xlarge}}                                                           & \textbf{\texttt{m1.large}}                                                            & \textbf{\texttt{m1.medium}}                                                           & \textbf{\texttt{m1.small}}                                                      \\ \hline
    \textit{Fire}                                   & \begin{tabular}[c]{@{}l@{}}[1032.11,\\1341.75]\end{tabular}   & \begin{tabular}[c]{@{}l@{}}[1010.62,\\1303.02]\end{tabular}   & \begin{tabular}[c]{@{}l@{}}[964.76,\\1259.68]\end{tabular}   & \begin{tabular}[c]{@{}l@{}}[670.82,\\872.85]\end{tabular}   \\ \hline
    \textit{HAR}                       & \begin{tabular}[c]{@{}l@{}}[88.27,\\ 94.20]\end{tabular}   & \begin{tabular}[c]{@{}l@{}}[99.84,\\ 104.49]\end{tabular}   & \begin{tabular}[c]{@{}l@{}}[122.33,\\ 126.67]\end{tabular}   & \begin{tabular}[c]{@{}l@{}}[135.13,\\ 137.92]\end{tabular}   \\ \hline
    \textit{Oil}          & \begin{tabular}[c]{@{}l@{}}[18083.59,\\18628.64]\end{tabular}   & \begin{tabular}[c]{@{}l@{}}[11159.71,\\11662.41]\end{tabular}   & \begin{tabular}[c]{@{}l@{}}[6139.59,\\6262.15]\end{tabular}                                                               & N/A                                                                 \\ \hline
    \textit{AIE}        & \begin{tabular}[c]{@{}l@{}}[237710.12,\\252686.82]\end{tabular} & \begin{tabular}[c]{@{}l@{}}[247166.73,\\251673.68]\end{tabular} & \begin{tabular}[c]{@{}l@{}}[168804.58,\\268273.11]\end{tabular} & \begin{tabular}[c]{@{}l@{}}[199676.71,\\203681.17]\end{tabular} \\ \hline
\end{tabular}}
\caption{\small{Confidence intervals of MIPS values for different DNN-based applications in Chameleon machines, resulted from Jackknife re-sampling method.}}
\label{jackCham}
\end{table}
\subsubsection{Estimating Confidence Interval using Jackknife Method} Let $p$ be the number of observed inference times. The Jackknife method calculates the harmonic mean in $p$ iterations, each time by eliminating one sample. That is, each time it creates a new sample (re-sample) with size $p-1$. Let $x_j$ be the $j$th observed inference time. Then, the harmonic mean of re-sample $i$ is called the pseudo-harmonic value (denoted as $y_i$) and is calculated based on Equation~\ref{jackEqu}.
\begin{equation}
    y_{i}=\frac{p-1}{\sum\limits_{j=1,j\neq i}^{p}\frac{1}{x_{j}}}
    \label{jackEqu}
\end{equation}
Next, the arithmetic mean (denoted $\bar{y}$) of the $p$ pseudo-harmonic values is computed, and is used to estimate the standard deviation. Finally, the t-distribution table is used to calculate the CI boundaries with a 95\% confidence level. The result of the Jackknife method for AWS machines is shown in Table \ref{jackAWS} that conforms with the MIPS calculation in Table \ref{mipsAWS}. Similarly, the results of analysis for Chameleon cloud using Jackknife method, shown in Table \ref{jackCham}, validate the prior MIPS calculations in Table \ref{mipsCham}. 
However, in the next part, we cross-validate these results using Bootstrap method.
\subsubsection{Estimating Confidence Interval using Bootstrap Method} Bootstrap repeatedly performs random sampling with a replacement technique \cite{patil2010using} on the observed inference times. The random sampling refers to the selection of a sample with the chance of non-zero probability and the number (represented as $k$) of re-sample data depends on the user's consideration. After re-sampling, the harmonic means of $k$ number of samples are calculated and sorted in ascending order to estimate the confidence intervals. Finally, for a specific confidence level, the ($\alpha / 2 \times k$)th and ($(1 - \alpha / 2) \times k$)th values are selected from the sorted samples as the lower and upper bounds of the CI. We set the $k$ value to 100 and $\alpha$ to 0.05 for 95\% confidence level.\looseness=-1   

For both AWS and Chameleon, the results of CI analysis using the Bootstrap method are similar to, thus validate, the ranges estimated by the Jackknife method. Therefore, due to the shortage of space, we do not report the table of MIPS values for the Bootstrap method. However, we note that the CI ranges provided by the Bootstrap method are shorter (\ie have less uncertainty), regardless of the application type and the cloud platform. The reason for the shorter range is that Bootstrap performs re-sampling with a user-defined number of samples that can be larger than the original sample size.

\begin{figure}[h!]
	\centering	
	\includegraphics[scale=0.60]{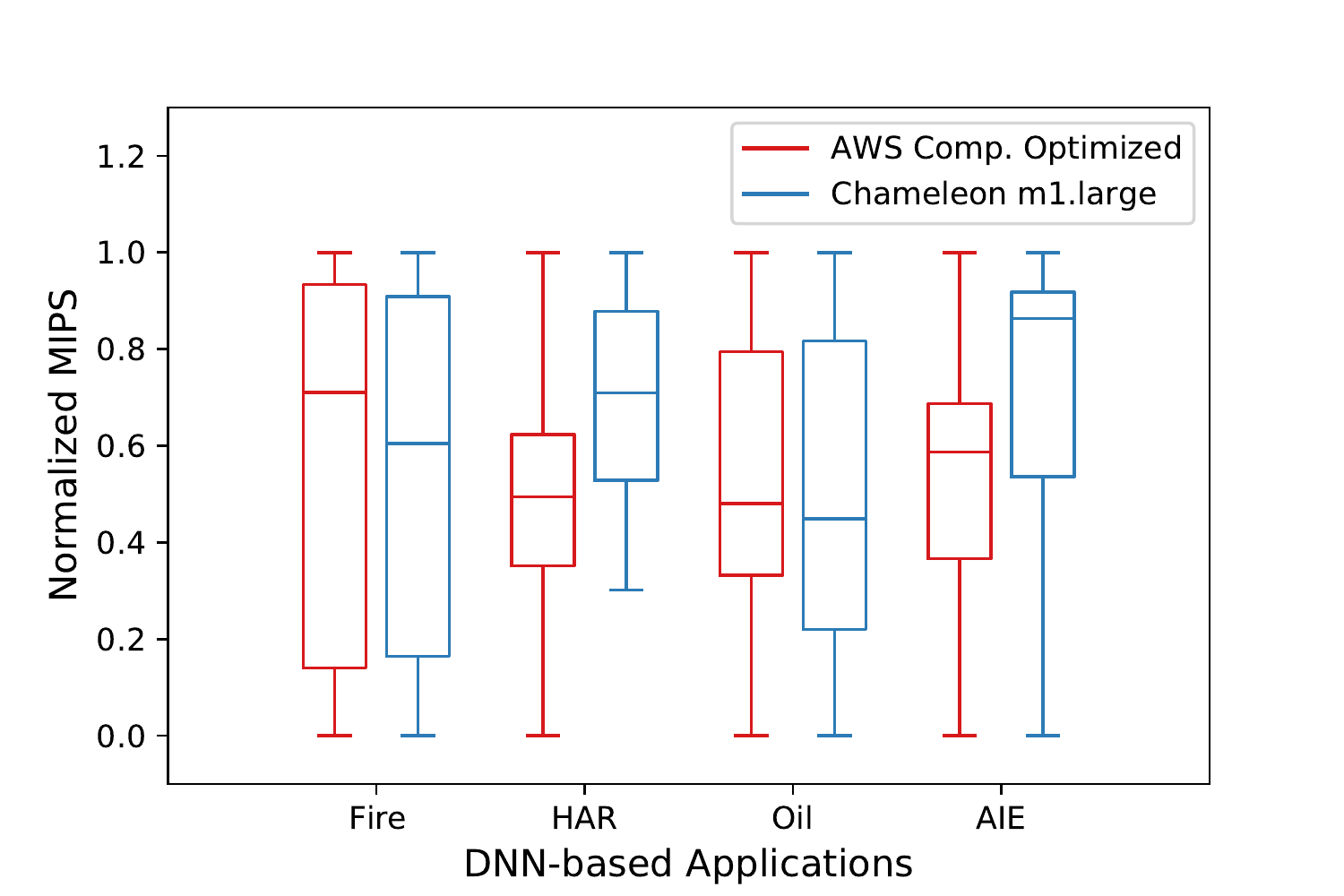}
	\caption{\small{Comparative analysis of the MIPS values of AWS and Chameleon machines for various DNN-based applications. For the sake of presentation, the MIPS values are  normalized between [0,1].}\label{fig:boxplot}}
\end{figure}

To perform a cross-platform analysis of the MIPS values, in Figure~\ref{fig:boxplot}, we compare the range of MIPS values for AWS \texttt{Compute Optimized} against \texttt{m1.large} that is a compatible machine type in Chameleon (see Tables~\ref{awsInstances} and~\ref{tab:chameleon}). The horizontal axis of this figure shows different application types and the vertical axis shows the MIPS values, normalized based on MinMax Scaling  in the range of [0,1], for the sake of better presentation.
Due to high variation in the input videos, we observe a broad CI range for Fire detection across both cloud platforms. 
However, for HAR, Oil Spill, and AIE applications, we observe that the first and third quartiles of the CI range in Chameleon (whose machines are prone to more transient failures \cite{charyyev2019towards}) is larger than those in AWS. This wide range indicates that, apart from variations in the input data, the reliability of underlying resources is also decisive on the stochasticity of the inference times. 



\section{Related Work}\label{relatedwork}
The advent of Industry 4.0 brought revolution in the O\&G industry, and Oil and Gas 4.0 \cite{lu2019oil} era arises. To deploy and improve the existing industrial DNN-based solutions for Oil and Gas 4.0, performance modeling of DNN-based applications is of great importance for cloud solution architects and researchers. Although, it is hard to find in-depth performance study of real-world DNN-based applications in cloud platforms, especially in the context of the O\&G industry considering the inference time of DNN-based applications. As such, we explore research areas related to benchmarking of DNN-based applications in various contexts (\ie artificial intelligence, video streaming, image recognition) utilizing local machines, virtual machines, or even mobile devices. However, these works focus mainly on training times, whereas our work focuses on the inference time.

As an execution platform for video transcoding operation, heterogeneous cloud virtual machines are analyzed for performance evaluation of different video content by Li \etal in \cite{li2018performance}. In this work, the authors elaborately discuss the correlation of video content type with the transcoding operation's execution time on heterogeneous VMs. In a similar context, Ghatrehsamani \etal in \cite{cpu-pinning2020} performed a detailed study of performance overhead of various cloud execution platforms utilizing four categories of real-world applications. One of the interesting findings of this work is that the containers are not always suitable computing platforms in the cloud. The above mentioned works mainly focus on virtual computing platforms regardless of application domain whereas we explore specific DNN-based applications inference execution pattern.

To perform the layer-wise behavior of various DNN models in heterogeneous deployment platforms, Xia \etal in \cite{xia2019dnntune} proposed a DNN tuning framework. This work provides a detailed discussion of the performance and energy consumption of DNN models (\ie CNN, LSTM, MLP) concerning different deployment strategies (\ie cloud, mobile device, mobile-cloud hybrid). Similarly, Turner \etal in \cite{turner2018characterising} proposed various mainstream DNN compression methods with the evaluation of the model accuracy, training time, and memory on two types of execution hardware (\ie CPU, GPU). In \cite{bianco2018benchmark}, Bianco \etal performed an extensive system-level analysis of a wide range of DNN models on two different computing architectures, namely NVIDIA Titan X Pascal and NVIDIA Jetson TX1.
~From an intuitively different perspective, the authors of \cite{richins2020ai} represent a detailed analysis of DNN-based application execution in optimized hardware and their overhead considering various prior and post operations during implementation on a special-purpose accelerator. This work proposes to utilize a specialized edge data center designed for DNN-based application that overcomes the data processing overhead. On the contrary, our work focuses on inference time behavior of DNN-based applications on heterogeneous cloud machine types, especially in the smart O\&G industry.


\section{Summary and Discussion}\label{concl}
Accurately estimating the inference time of latency-sensitive DNN-based applications plays a critical role in robustness and safety of Industry 4.0. Such accurate estimations enable cloud providers and solution architects to devise resource allocation and load balancing solutions that are robust against uncertainty exists in the execution time of DNN-based applications. In this work, we provide application- and resource-centric analyses on the uncertainty exists in the inference times of several DNN-based applications deployed on heterogeneous machines of two cloud platforms, namely AWS and Chameleon. In the first part, we utilized the Shapiro-Wilk test to verify if the assumption of Normal distribution for the inference time holds. We observed that the inference times often do not follow a Normal distribution. Therefore, in the second part, we broaden our distribution testing investigation and utilized the Kolmogorov-Smirnov test to verify the underlying distributions in each case. The analysis showed that inference times across the two cloud platforms often follow Student's t-distribution. However, in several cases in Chameleon cloud we observed the Log-normal distribution that we attribute it to the uncertain performance of VMs in this platform. 
Next, to conduct a resource-centric analysis, we modeled MIPS (as a rate-based performance metric) of the heterogeneous machines for each application type. In the analysis, we took a non-parametric approach, which is suitable for rate-based metrics, and utilized the Jackknife and Bootstrap re-sampling methods with harmonic mean to determine the range of confidence intervals of the MIPS values in each case. The calculated MIPS values and their CI ranges reflect the behavior of different DNN-based applications under various machine types and cloud platforms. A comparative analysis of the CI ranges across AWS and Chameleon cloud demonstrate that the uncertainty in the inference time is because of variations in the input data and unreliability of the underlying platforms. 
In the future, we plan to incorporate the findings of this research to devise accurate resource allocation methods in IoT and edge computing systems. In addition, we plan to develop a predictive analysis to determine the execution of each inference task upon arrival.      

\section*{Acknowledgments}
The research was supported by Chameleon cloud and Amazon
Cloud (AWS) research credit.
%
\linespread{0.90}
\bibliographystyle{ieeetr}
\bibliography{references}


\end{document}